# On a possibility of effective electron attraction without "a glue"

Valentin Voroshilov

Physics Department, Boston University, Boston, MA, 02215, USA

With the means of Schrödinger equation a two electron problem is considered. It is shown that electrons in a periodic potential might experience effective attraction without any mediating agents; which means electrons in HTSC might have effective attraction potential.

Despite being under attack for almost twenty-five years there is no yet a commonly accepted solution for the problem of high temperature superconductivity[1]. Among different approaches there are theoretical[2] and experimental[3] views propelling the idea that electrons in cuprates do not need any mediating agents to experience effective attraction. If that is a case, one would expect that a two electron problem would have solutions corresponded to effective attraction between the electrons. To examine this hypothesis we write Schrödinger equation (1) for two electrons interacting with each other and with the lattice and the rest of the electron system

$$\left[ -\frac{1}{2}\Delta_1 - \frac{1}{2}\Delta_2 + U_1 + U_2 + V_{12} \right] \Psi_{12} = E\Psi_{12}. \qquad (1)$$

In (1) Planck's constant and electron mass are set to unity; $\Psi_{12}$ is a space part of an antisymmetric wave function for two electrons; $U_i$ describes interaction between $i$-th electron ($i = 1,2.$) and the lattice and the rest of the electron system ($N - 2$ electrons); $V_{12}$ describes interaction between the two electrons. We assume that $U_i(\vec{r}+\vec{R})=U_i(\vec{r})$ where $\vec{R}$ is any of the lattice vectors (this assumption is based on the assumption that electron density of $N-2$ electrons is periodic or almost periodic in space and resembles periodic properties of the lattice); and $V_{12}=V(|\vec{r}_1-\vec{r}_2|)$. The simplest way to account for interaction between the two electrons and the rest of the electron system is by stating the requirement that the two electrons cannot occupy individual states occupied by all other electrons[4]. If interaction term $V_{12}$ were small one could use a perturbation theory. However a strong Coulomb repulsion makes this problem difficult to analyze. The idea is to replace the actual interactions with some effective ones, which would allow application of a perturbation theory.

Let us rewrite equation (1) as following

$$\left[-\frac{1}{2}\Delta_1-\frac{1}{2}\Delta_2+U_{1eff}+U_{2eff}+W_{12}\right]\Psi_{12}=E\Psi_{12}. \tag{2}$$

In (2) we assume $W_{12}$ is small, and $U_{i\,eff}(\vec{r}_i+\vec{R})=U_{i\,eff}(\vec{r}_i)$. In that case perturbation theory allows us to write

$$E=\varepsilon_{\vec{p}_1}+\varepsilon_{\vec{p}_2}+E^{(1)}+E^{(2)}+...;\qquad \Psi_{12}\approx\Phi_{12}=\frac{1}{\sqrt{2}}\{\psi_{\vec{p}_1}(\vec{r}_1)\psi_{\vec{p}_2}(\vec{r}_2)+\sigma\psi_{\vec{p}_2}(\vec{r}_1)\psi_{\vec{p}_1}(\vec{r}_2)\}. \tag{3}$$

In (3) $\sigma=\pm 1$ (depending on the spin configuration); and $\psi_{\vec{p}}(\vec{r})$ represents an orthogonal set of normalized Bloch or Wannier wave functions satisfying equation (4)

$$\left[-\frac{1}{2}\Delta+U_{eff}\right]\psi_{\vec{p}}(\vec{r})=\varepsilon_{\vec{p}}\psi_{\vec{p}}(\vec{r}). \tag{4}$$

The success of the calculation is heavily based on the choice of effective term $U_{eff}$. Since a strong Coulomb repulsion plays the major role, we are looking for a way to "extract" it from term $V_{12}$ and include it in terms $U_{i\,eff}$. In the spirit of Hartree-Fock approximation we could write

$$V_{12} \approx \int V(|\vec{r}_1 - \vec{r}_2|)\rho(\vec{r}_2)d\vec{r} = \frac{1}{2}\int V(|\vec{r}_1 - \vec{r}_2|)\rho(\vec{r}_2)d\vec{r}_2 + \frac{1}{2}\int V(|\vec{r}_1 - \vec{r}_2|)\rho(\vec{r}_1)d\vec{r}_1; \quad (5a)$$

$$\text{with} \quad \rho(\vec{r}) = \int |\Psi(\vec{r}_1,\vec{r})|^2 d\vec{r}_1 = \int |\Psi(\vec{r},\vec{r}_2)|^2 d\vec{r}_2. \quad (5b)$$

This gives us an idea on how to include Coulomb repulsion into terms $U_{i\,eff}$. To achieve this goal let us write the effective interaction terms as the following

$$U_{i\,eff}(\vec{r}_i) = U_i(\vec{r}_i) + C_i(\vec{r}_i); \quad C_i(\vec{r}_i) = \frac{1}{2}\int V(|\vec{r}_j - \vec{r}_i|)\rho(\vec{r}_j)d\vec{r}_j. \quad (6)$$

The effective interaction between the two electrons can be written now in the following form

$$W(\vec{r}_1,\vec{r}_2) = V(|\vec{r}_1 - \vec{r}_2|) - C(\vec{r}_1) - C(\vec{r}_2). \quad (7)$$

According to our assumption effective interaction term (7) is small enough to apply a perturbation theory. If we use an approximate expression (3) for wave function $\Psi_{12}$, expression (5) gives

$$\rho(\vec{r}) \approx \int |\Phi(\vec{r}_1,\vec{r})|^2 d\vec{r}_1 = \frac{1}{2}(1 + \sigma\delta_{\vec{p}_1\vec{p}_2})\{|\psi_{\vec{p}_1}(\vec{r})|^2 + |\psi_{\vec{p}_2}(\vec{r})|^2\} = \frac{|\psi_{\vec{p}_1}(\vec{r})|^2 + |\psi_{\vec{p}_2}(\vec{r})|^2}{2}. \quad (8)$$

In (8) we make a natural assumption that paired electrons have different momenta. Expressions (4), (6) and (8) are self-consistent in terms of periodic properties, i.e. if $\rho(\vec{r})$ is periodic, so are $C(\vec{r})$ and $U_{i\,eff}(\vec{r}_i)$, and vice versa.

The first correction to the energy $E^{(1)}$ is provided by a standard expression (9)

$$E^{(1)} = \int \Phi_{12}^* W_{12} \Phi_{12} d\vec{r}_1 d\vec{r}_2. \quad (9)$$

If electrons are mostly localized (which is also a natural assumption for electrons in a narrow energy band) we assume that only on-site repulsion plays the major role. In a spirit of a Hubbard model we replace the exact Coulomb interaction with a Dirac delta-function, $V_{12} = \lambda\delta(\vec{r}_1 - \vec{r}_2)$ ($\lambda > 0$). A calculation of the first correction to the energy gives in that case

$$E^{(1)} = \frac{\lambda}{2}\left[(1+2\sigma)\int |\psi_{\vec{p}_1}(\vec{r})|^2 |\psi_{\vec{p}_2}(\vec{r})|^2 d\vec{r} - \frac{1}{2}\int\{|\psi_{\vec{p}_1}(\vec{r})|^4 + |\psi_{\vec{p}_2}(\vec{r})|^4\}d\vec{r}\right]. \tag{10}$$

Expression (10) leads to a conclusion that if electrons have parallel spins ($\sigma = -1$), the first correction to the interaction energy is negative, which means such electrons are effectively attracting each other.

The perturbation theory is applicable only if effective interaction $W_{12}$ is small, i.e. if (for all allowed states)

$$|(W_{12})_{\vec{p}_1\vec{p}_2\vec{q}_1\vec{q}_2}| << |\varepsilon_{\vec{p}_1} + \varepsilon_{\vec{p}_2} - \varepsilon_{\vec{q}_1} - \varepsilon_{\vec{q}_2}|. \tag{11}$$

Expression (11) means that, when calculated for all allowed states, values of $\Delta\varepsilon = |\varepsilon_{\vec{p}_1} + \varepsilon_{\vec{p}_2} - \varepsilon_{\vec{q}_1} - \varepsilon_{\vec{q}_2}|$ should be at least limited from below by some positive number $0 < \Delta\varepsilon_{\lim} < \Delta\varepsilon$.

For free particles $\varepsilon_{\vec{p}} = \frac{p^2}{2}$, hence limit $0 < \Delta\varepsilon_{\lim}$ does not exist (there are states with $\Delta\varepsilon = 0$).

As long as both terms $U_{i\,eff}(\vec{r}_i)$ are periodic, for a narrow energy band Eq. (4) gives spectrum $\varepsilon_{\vec{p}}$ corresponded to the symmetries of the lattice, for example for a square 2D lattice (with $R_x = R_y = 1$) energy spectrum is $\varepsilon_{\vec{p}} \approx -t(\cos(p_x) + \cos(p_y))$. If we set $\vec{p}_1 = -\vec{p}_2$ we have $\Delta\varepsilon = |2\varepsilon_{\vec{p}_1} - \varepsilon_{\vec{q}_1} - \varepsilon_{\vec{q}_2}|$ and limit $0 < \Delta\varepsilon_{\lim}$ does not exist.

There is however another possibility, when $\varepsilon_{\vec{p}_1} + \varepsilon_{\vec{p}_2} = 0$ and all electron states below a certain positive energy level $\varepsilon_{\max}$ are occupied ($0 < \varepsilon_{\max} < 2t$; without a further analysis we should not call $\varepsilon_{\max}$ a Fermi level). In that case $\Delta\varepsilon_{\lim} = 2\varepsilon_{\max}$; and the second correction to the energy of the two electrons is negative, too. A similar idea was used for introducing a model Hamiltonian for a system of electrons with effective attraction between electrons having the same group velocity[5]; now there is a theoretical base for this view
($\frac{\partial \varepsilon_{\vec{p}_1}}{\partial \vec{p}_1} = \frac{\partial \varepsilon_{\vec{p}_1}}{\partial \vec{p}_1}$ ; => $\sin p_{1x,y} = \sin p_{2x,y}$ ; => $\cos p_{1x,y} = -\cos p_{2x,y}$ ; => $\varepsilon_{\vec{p}_1} + \varepsilon_{\vec{p}_2} = 0$ ).

The offered analysis supports the notion that electrons in a periodic potential can experience effective attraction without any mediating agents. Further investigation is needed to analyze how a deviation from the assumptions made above would affect the conclusion.

---

[1] J.G. Bednorz and K.A. Muller, Z. Physik **B 64**, 189 (1986).

[2] Philip W. Anderson: Do We Need (or Want) a Bosonic Glue to Pair Electrons in High Tc Superconductors?; http://arxiv.org/abs/cond-mat/0609040; Philip W. Anderson: Is There Glue in Cuprate Superconductors? http://www.sciencemag.org/cgi/content/full/316/5832/1705

[3] Where's The Glue? Scientists Find A Surprise When They Look For What Binds In Superconductivity: http://www.sciencedaily.com/releases/2008/04/080410140538.htm

[4] Leon N Coper; Bound Electron Pairs in a Degenerate Fermi Gas // Phys. Rev. Lett., **V 104**, #4, November (1956), pp. 1189-1190

[5] On electron pairing in a periodic potential // Physica C: Supercunductivity, **V 470**, # 21, November (2010), pp. 1962 – 1963 // http://dx.doi.org/10.1016/j.physc.2010.08.007